\documentclass[]{aa}
\usepackage{epsfig}

\newcommand{\BS}{{{\it Beppo}SAX}\ \ignorespaces}
\newcommand{\ltsima} {$\; \buildrel < \over \sim \;$}
\newcommand{\gtsima} {$\; \buildrel > \over \sim \;$}
\newcommand{\lta} {\lower.5ex\hbox{\ltsima}}
\newcommand{\gta} {\lower.5ex\hbox{\gtsima}}

\newcommand{\phd} {\phantom{\# }}

\begin{document}
\thesaurus{11.17.4 3C 273; 13.25.2}

\title{The hidden X--ray Seyfert nucleus in 3C~273: {\it Beppo}SAX Results}

\author{
     F.~Haardt\inst{1}
\and G.~Fossati\inst{2} 
\and P.~Grandi\inst{3}
\and A.~Celotti\inst{2} 
\and E.~Pian\inst{4}
\and G.~Ghisellini\inst{5}
\and A.~Malizia\inst{6,7}
\and L.~Maraschi\inst{5}
\and W.~Paciesas\inst{8} 
\and C.M.~Raiteri\inst{9}
\and G.~Tagliaferri\inst{5}
\and A.~Treves\inst{10}
\and C.M.~Urry\inst{11} 
\and M.~Villata\inst{9}
\and S.~Wagner\inst{12} 
}

% \and L.~Bassani\inst{4}
 % \and M.~Cappi\inst{4}
% \and F.~Frontera\inst{4}
% \and G.~Malaguti\inst{4}
 % \and L.~Nicastro\inst{4}
% \and L.~Chiappetti\inst{7}
 % \and S.~Molendi\inst{7}
 % \and A.~Comastri\inst{8}
% \and S.~Giarrusso\inst{9}
 % \and G.~Palumbo\inst{10}
% \and C.~Perola\inst{11}
 % \and M.~Salvati\inst{13}
% \and R.~Fanti\inst{14}
% \and R.~Mantovani\inst{14}
 % \and T.~Venturi\inst{14}
 % \and E.~Massaro\inst{15}
% \and R.~Falomo\inst{16}
% \and W.~Brinkmann\inst{17}
 % \and M.~Guainazzi\inst{18}
% \and R.~Sambruna\inst{20} 
 % \and M.~Sikora\inst{21} 
% \and R.~Staubert\inst{22}
 % \and P.~Padovani\inst{23} 

\institute{
Dipartimento di Fisica dell'Universit\`a di Milano, Milano, Italy
\and
International School for Advanced Studies, SISSA/ISAS, Trieste, Italy
\and
IAS, CNR, Frascati (RM), Italy
\and
ITESRE, CNR, Bologna, Italy
\and
Osservatorio Astronomico di Brera/Merate, Milano/Lecco, Italy
\and
Department of Physics \& Astronomy, University of Southampton, UK 
\and
\BS Science Data Center, Roma, Italy
\and
University of Alabama in Huntsville, Huntsville, AL, USA
\and
Osservatorio Astronomico di Torino, Pino Torinese (TO), Italy
\and
Dipartimento di Fisica dell'Universit\`a di Como, Como, Italy
\and
STScI, Baltimore, MD, USA
\and
Landessternwarte, Heidelberg, Germany 
}

%IFCTR, CNR, Milano, Italy
%Osservatorio Astronomico, Bologna, Italy
%IFCAI, CNR, Palermo, Italy
%Dipartimento di Astronomia dell'Universit\`a di Bologna, Bologna, Italy
%Dipartimento di Fisica della III Universit\`a di Roma, Roma, Italy
%Osservatorio Astrofisico di Arcetri, Firenze, Italy
%IRA, CNR, Bologna, Italy
%Istituto Astronomico, Universit\`a di Roma ``La Sapienza", Roma, Italy
%Osservatorio Astronomico di Padova, Padova, Italy
%MPE, Garching, Germany
%ESTEC/SA, ESA, Noordwijk, The Netherlands
%Penn State University, State College, PA, USA
%Copernicus Astronomical Center, Warsaw, Poland
%University of Tubingen, Germany

\offprints{haardt@uni.mi.astro.it}

\date{Received ; accepted}

\maketitle

\markboth{F. Haardt et al.} {The hidden Seyfert in 3C 273: \BS observations}  
                          
\begin{abstract}
We present the results of 5 \BS AO1 Core Program observations of
3C~273 performed in Jan. 1997 and compare them in detail with data
obtained during the satellite Science Verification Phase (SVP), in Jul. 
1996 (Grandi et al. 1997).

3C 273 was about 15\% brighter in the first 1997 observation than in
the last one, and, on average, a factor 2 brighter than the SVP
observation.  A count rate variation in the 2--10 keV band of $\simeq
12$\% in $\sim 0.5$ day was clearly detected during the last of the four
pointings. 

Power--law fits with Galactic absorption to all observations yield
spectral indices in the range $\Gamma=1.53-1.6$. Though a power
law is an acceptable representation of the data in the whole
0.1--200 keV range, there is indication of a steepening of the
spectrum as the energy increases.  Residuals with respect to a single
power law suggest the presence of a weak fluorescence iron line in the
MECS data. No other features are detectable. 
Our data therefore mark a difference with respect to the SVP
data, where a steeper power law below 0.5 keV, an absorption feature at
$\sim 0.6$ keV, and a more prominent fluorescence iron line have been
found.

The weakening of cold/warm matter signatures in our data with respect
to the SVP ones may indicate that, at higher luminosities, the
featureless continuum produced in a relativistic jet overwhelms any
thermal and/or reprocessed radiation, while the two components were almost 
comparable during the lower state of Jul. 1996.

We quantitatively test this scenario, by considering an emission model
which comprises the contribution from a thermal Seyfert--like nucleus
and a non--thermal power--law component, and find that indeed the
observed features in both the AO1 and SVP data are
consistently reproduced by varying only the intensity of the
non--thermal emission. Within this scenario, this radio--loud source
shows evidence not only for thermal disk--like emission, but also
substantial reprocessing of X--rays onto cold matter.  There is no
evidence of a direct correlation between the two components.


\end{abstract}
 
\section{Introduction}

3C 273 is a nearby ($z=0.158$) quasar, and is one of the extragalactic
objects best studied across the entire electromagnetic spectrum. It
shows almost all the features proper of high--luminosity quasars,
i.e. radio and optical jet with high polarization, double radio lobes,
superluminal motion, variability at all frequencies, broad emission
lines and signs of thermal emission in the UV band ({\it Big Blue
Bump}).  Multifrequency campaigns (e.g. Courvoisier et al. 1987) 
showed that the broad band Spectral
Energy Distribution (SED) is rather complex, with two clear peaks at
UV ($\sim 10$ eV) and $\gamma-$ray ($\sim 1-10$ MeV) energies (Ramos et
al. 1997, von Montigny et al. 1997). The general interpretation is
that the radio--to--$\gamma-$ray continuum is produced by the jet
either via synchrotron and inverse Compton emission (e.g.  Marscher \&
Travis 1996, Ghisellini et al. 1996), or by a proton-initiated cascade
(Mannheim 1993), while the accretion disk is responsible for the UV
bump (see von Montigny et al. 1997, and discussion therein). Part of
the accretion disk thermal emission could be Comptonized in an accretion disk 
corona (e.g., Haardt \& Maraschi 1991), or in the hot
two temperature inner disk region, along the line originally proposed
by Shapiro, Lightman \& Eardly (1976), and more recently by
Chakrabarty \& Titarchuk (1995).

The first detection in the X--ray band was obtained by Bowyer et
al. (1970).  Observations in the medium--hard X--ray range (up to 30
keV with {\it GINGA}) showed a hard power law continuum with photon index
ranging between $1.3-1.6$ (Turner et al. 1990, 1991; Williams et
al. 1992; Yaqoob et al. 1994), while both EXOSAT (Turner et
al. 1995) and, more recently, {\it ROSAT} data (Staubert 1992) reveal
evidence of an excess above the extrapolation of the hard power law at
energies $\lta 1$ keV. The study of the spectrum as seen by {\it
ROSAT} (Leach, McHardy \& Papadakis 1995) indicates the presence of a
soft component (0.1--0.3 keV) with photon index $\Gamma\simeq 2.7$
uncorrelated with the hard X--ray power law. Cold matter signatures
such as a Compton reflection hump and an iron line were barely
detected by {\it GINGA} (Williams et al. 1992). An emission line at
6.4 keV (source frame) seen at low flux levels was later confirmed by
{\it ASCA} (Cappi \& Matsuoka 1996).

3C 273 shows prominent $\gamma-$ray emission as well, detected by OSSE
(Johnson et al. 1995), COMPTEL (Hermsen et al. 1993) and EGRET
(Hartman et al. 1992), on board the {\it Compton Gamma--Ray
Observatory} ({\it CGRO}). OSSE data reveal that the hard power law
extends up to $\sim$ 1 MeV, with a break at higher energies
(McNaron--Brown et al. 1995).  

Recently (Jul. 1996), the \BS SVP observation of 3C 273 has, for the first time,
made available a simultaneous coverage of the 0.2--200 keV
energy band. Grandi et al. (1997) confirmed that the
$\Gamma\simeq 1.5$ power law extends at least up to 200 keV without
any appreciable deviation, neither a Compton reflection hump nor a
break. At energies below 1 keV both an absorption feature {\it and} a
soft excess were present in the spectrum. A weak iron line at 6.4 keV
(source frame) was also present, with an equivalent width (EW) 
$\simeq$ 30 eV, consistent with {\it GINGA} (Williams et al. 1992) and 
ASCA (Cappi \& Matsuoka 1996) results.

Here we present and discuss the results of pointings of 3C 273 performed in
Jan. 1997. We will also make a comparison of these data with the
SVP data, in view of understanding the long term behaviour of the
source. The 3C 273 observations discussed here are part of the AO1 Core
Program dedicated to bright blazars. 

The paper is organized as follows: in \S2 observations and data reduction
techniques are described, while \S3 is devoted to the presentation of data
analysis procedures. In \S4 we present BATSE and optical data obtained during simultaneous observations. Results are then discussed, together with implications for
theoretical models, in \S5. Finally, a brief summary and conclusions are drawn
in \S6. Results of a preliminary analysis have been discussed in Maraschi, 
Fossati \& Haardt (1997) and Haardt et al. (1997). If not otherwise indicated,
hereinafter energies are measured in the observer frame. 

\section{Observations and data reduction}

%---------------------------TABLE 1----------------------------------------
\begin{table*}
\caption[]{Observation Log\label{tab:log}}
\begin{tabular}{lccccccc}
\noalign{\hrule}
\multicolumn{1} {l}{Obs. (start date)}
&\multicolumn{3}{c}{Net Exposure Time (ks)}
&~~~
&\multicolumn{3}{c}{Count rate (cts/s)}\\
&&&&&&&\\
&\multicolumn{1}{c}{LECS}
&\multicolumn{1}{c}{MECS}
&\multicolumn{1}{c}{PDS$^{\rm a}$}
&
&\multicolumn{1}{c}{LECS$^{\rm b}$}
&\multicolumn{1}{c}{MECS$^{\rm c}$}
&\multicolumn{1}{c}{PDS$^{\rm d}$}\\
\noalign{\hrule}
&&&&&&&\\
{A (01/13/97)} &
{ 13.6 } & { 25.4 } & { 22.8 } && {$ 0.802 \pm 0.008 $} &
{$ 1.995 \pm 0.009 $} & {$ 1.40 \pm 0.04 $} \\
&&&&&&&\\
{B (01/15/97)} &
{ 13.3 } & { 24.1 } & { 21.6 } && {$ 0.729 \pm 0.007 $} &
{$ 1.883 \pm 0.009 $} & {$ 1.30 \pm 0.04 $} \\
&&&&&&&\\
{C (01/17/97)} &
{ 12.5 } & { 27.6 } & { 25.0 } && {$ 0.698 \pm 0.008 $} &
{$ 1.795 \pm 0.008 $} & {$ 1.24 \pm 0.04 $} \\
&&&&&&&\\
{D (01/22/97)} &
{ \phantom{8}8.8 } & { 22.5 } & { 18.6 } && {$ 0.695 \pm 0.009 $} &
{$ 1.691 \pm 0.009 $} & {$ 1.20 \pm 0.04 $} \\
&&&&&&&\\
\noalign{\hrule}
\multicolumn{8}{l}{$^{\rm a}$ Exposure time for half of the PDS effective area.}\\
\multicolumn{8}{l}{$^{\rm b}$ 0.12--4.0 keV count rate.}\\
\multicolumn{8}{l}{$^{\rm c}$ 1.6--10.5 keV count rate.}\\
\multicolumn{8}{l}{$^{\rm d}$ 13--200 keV count rate on half of the effective 
area.}\\
\end{tabular}
\end{table*}
%-----------------------------------------------------------------------------

\subsection{The \BS mission in brief} 

For an exhaustive description
of the Italian/Dutch \BS mission we refer to Boella et
al. (1997). Here we briefly review the main characteristics of the
instrumentation, and the data reduction techniques applied.

The narrow field coaligned instrumentation on \BS consists in a Low Energy
Concentrator Spectrometer (LECS), three Medium Energy Concentrator
Spectrometers (MECS), a High Pressure Gas Scintillation Proportional Counter
(HPGSPC), and a Phoswich Detector System (PDS). The LECS
and MECS have imaging capabilities in the 0.1--10 keV and 1.3--10 keV energy
band, respectively, with energy resolution of $\simeq 8$\% at 6 keV. At the
same energy, the angular resolution is about 1.2 arcmin (Half Power Radius). In
the overlapping energy range the MECS effective area ($\simeq 150$ cm$^2$) is
$\sim 3$ times that of the LECS. The HPGSPC covers the range 4--120 keV, and
the PDS the range 13--300 keV. In the overlapping energy interval, 
the PDS is more
sensitive, while the HPGSPC has a better energy resolution. HPGSPC data will
not be discussed in the present paper. 

\subsection{The \BS observations} 

The observations of 3C 273 were
performed as part of the \BS AO1 Core Program.  The source was
observed between Jan. 13--23, 1997. In this period 3C
273 has been observed 5 times: Jan. 13, observation period (OP)
1492, hereinafter referred as observation "A"; Jan. 15, OP 1502,
observation "B"; Jan. 17--18, OP 1513, observation "C"; and
Jan. 22--23, OP 1561 and OP 1563, observation "D".  The datasets
of the last two OPs have been merged into a single one due to the
extremely short exposure time of OP 1561.  The total effective
exposure was 45.2 ks in the LECS, 92.1 ks in the MECS, and
88.0 ks in the PDS. A journal of observations is given in
Table~\ref{tab:log}. 
For comparison, the net exposure times for LECS, MECS and PDS in the
SVP were 12, 131 and 128 ks, respectively.

\subsection{LECS and MECS data reduction} 
%The LECS and MECS analysis 
%presented is based on the {\tt SAXDAS} linearized and cleaned event
%files, together with appropriate background event files, as produced
%at the \BS Science Data Center (SDC) \footnote{For details, see
%documentation at the URL http://www.sdc.asi.it/software/cookbook}.

Light curves and spectra for the LECS and the MECS have been
accumulated using the {\tt Xronos} and {\tt Xselect} tools ({\tt
Ftools v4.0}). 
We extracted the events from within a radius of 8' for the LECS and 6' for
the composite MECS image, consisting of the three co--added MECS images. 
These selected events were then used to construct the light
curves and to accumulate energy spectra for each pointing. Different
spectral binnings have been used and tested, such as the
pre--constructed channel grouping templates as provided by the \BS Science 
Data Center (SDC)\footnote{For details, see documentation at URLs
http://www.sdc.asi.it/software/cookbook/spectral.html and
ftp://www.sdc.asi.it/pub/sax/cal/responses/grouping/}, or
more standard groupings aimed to achieve a given minimum S/N across
the entire spectrum. Unless otherwise indicated, the results of the
spectral analysis will refer to spectra accumulated over the {\tt
grouping2} template released by the SDC.

LECS data have been considered only in the range 0.12--4 keV due to
calibration problems at higher energies (Guainazzi 1997). The average
count rate in the four 3C 273 pointings is 0.726 cts/s in the LECS,
and 1.766 cts/s in the MECS. Spectral analysis has been performed with
the {\tt XSPEC 10.00} package, using the response matrices released in
Sept. 1997.

\subsection{PDS data reduction} The PDS was operated in collimator 
rocking mode, with a pair of units pointing at the source and the other
pair pointing at the background, the two pairs switching on and off
source every 96 seconds.

Source visibility windows were selected following the criteria of no
Earth occultation and high voltage stability during the exposures. In
addition, the observations closest to the South Atlantic anomaly were
discarded from the analysis. From collimators positions the ON and OFF
time windows were also created and merged with the source visibility
window to create the final time windows on which the source+background
and background spectra were accumulated for each of the four PDS
units, using the {\tt XAS} software package, with an improved
rise--time correction. 
The source was detected above 3 $\sigma$ up to $\sim$ 
200 keV in pointings A,B,C, and only up to $\sim$ 150 keV in D. 
The grouped spectra from the four units were then coadded.

In the calibrated energy band ($\sim$ 12--200 keV), the spectra 
were binned for all pointings in 15 energy intervals. 
The fits were performed with the {\tt XSPEC 10.00} package using a response
matrix appropriate for the improved, channel-dependent, subtraction of
the background, for the correction for the dead layer of the PDS crystals,
and for the improved rise--time correction.
The average PDS count rate in the four pointings is 1.29 cts/s, on
half of the effective area.

\begin{figure}
\epsfig{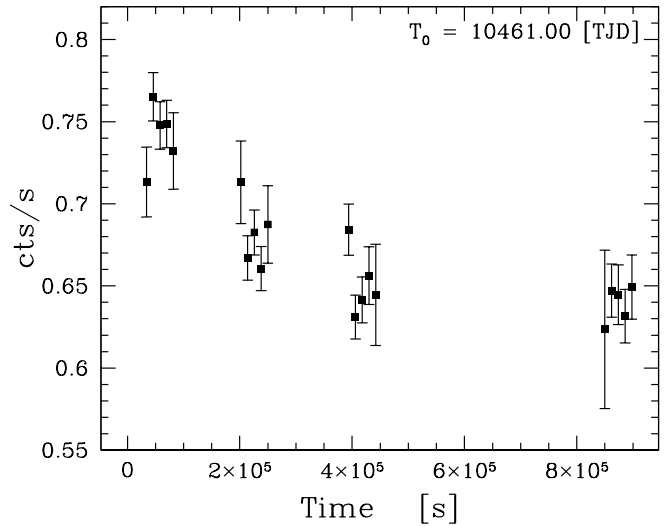} 
\vskip -1.5 cm
\caption{LECS light curve. The bin width adopted is 12 ks.
\label{fig:lc-lecs} }
\end{figure}

\section{\bf Data analysis and results}

\subsection{Temporal analysis} 

\begin{figure}
\epsfig{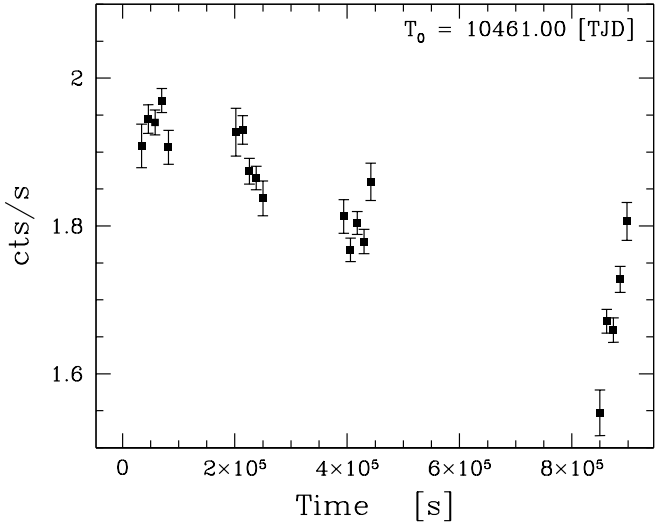}
\vskip -1.5 cm
\caption {MECS light curve. The bin width adopted is 12 ks.
\label{fig:lc-mecs} }
\end{figure}

\begin{figure}
\epsfig{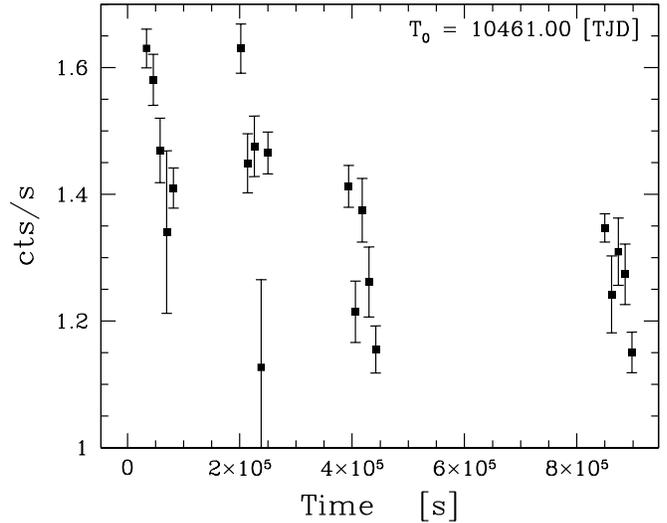}
\vskip -1.5 cm
\caption {PDS light curve. The bin width adopted is 12 ks.
\label{fig:lc-pds} }
\end{figure}

In Fig.~\ref{fig:lc-lecs}, \ref{fig:lc-mecs} and \ref{fig:lc-pds} 
the overall light curves for the three instruments are shown. A decreasing
trend is apparent in all the three instruments. We selected the MECS data
for a detailed temporal analysis, as they are the \BS data best 
suited for such a purpose, on account of better statistics and
reliable performance stability.

The MECS count rate monotonically
decreases on time scale of days (Fig.~\ref{fig:lc-mecs}).  
The difference in the
count rate between the first and last observation is $\simeq
15$\%. Similar variations are present also in the LECS and PDS data
(Table~\ref{tab:log}), although the paucity of data does not allow to simply
set limits on any lag between different spectral bands.
We applied the $\chi^2$ test to the total MECS light curve using time
bin size ranging from 100 to 8000 s, and found that, 
independently of the time binning used, the count rate variation is 
significant at 99.99\% level.

The general trend shows up also in the LECS and PDS (see 
Table~\ref{tab:log}, and Fig.~\ref{fig:lc-mecs} and~\ref{fig:lc-pds}), 
though the lower statistics does not allow a meaningful comparison with the 
MECS light curve.

We also checked for MECS variability within single pointings. 
While the first three observations are consistent with a
constant count rate, the last one is not, at the 99.88\% significance
level. The count rate reverses the decreasing trend and increases by
about $\sim 12$\% in about half a day.

%---------------------TABLE 2---------------------------------------------
\begin{table}
\caption[]{Power Law$^{\rm a}$ Fits: Broad Band\label{tab:broad-band-fits}}
\begin{tabular}{lcccc}
\noalign{\hrule}
\multicolumn{1}{l}{Obs.} &           
\multicolumn{1}{c}{$\Gamma$} &
\multicolumn{1}{c}{${\rm LE/ME}^{\rm b}$} &
\multicolumn{1}{c}{${\rm PDS/ME}^{\rm c}$} &
\multicolumn{1}{c}{$\chi^{2}/d.o.f$ ($\chi^{2}_\nu$)}\\
&&&&\\
\noalign{\hrule}
&&&&\\
{A} &  { $1.57_{-0.02}^{+0.02}$ } & 
{$0.76_{-0.03}^{+0.03}$} & { $0.86_{-0.06}^{+0.06}$ } & { \phantom{8}$63.2/84$ (0.75) }\\
&&&&\\ 
{B} &  { $1.58_{-0.02}^{+0.02}$ } &
{$0.70_{-0.03}^{+0.02}$} & { $0.94_{-0.07}^{+0.07}$ } & { $117.4/84$ (1.41) }\\
&&&&\\
{C} &  { $1.62_{-0.02}^{+0.03}$ } &
{$0.69_{-0.02}^{+0.03}$} & { $0.90_{-0.06}^{+0.07}$ } & { \phantom{8}$92.4/83$ (1.11) }\\
&&&&\\
{D} &  { $1.56_{-0.03}^{+0.02}$ } &
{$0.76_{-0.04}^{+0.03}$} & { $0.83_{-0.07}^{+0.08}$ } & { \phantom{8}$99.0/83$ (1.19) }\\
&&&&\\
\noalign{\hrule}
\end{tabular}
{\raggedright\par
{\bf Note:} quoted errors are 90\% confidence 
intervals for 3 interesting parameters ($\Delta\chi^2 = 6.25$).\par\noindent 
$^{\rm (a)}$ N$_{\rm H}$ fixed at Galactic value $1.69\times 
10^{20}$ cm$^{-2}$. \par\noindent
$^{\rm (b)}$ LECS power law normalization relative to the MECS. \par\noindent
$^{\rm (c)}$ PDS power law normalization relative to the MECS. \par }
\end{table}
%------------------------------------------------------------------------

		%---------------------------%

\subsection{Spectral analysis} 

We first present the broad band spectrum analysis and then 
consider each instrument separately.

		%---------------------------%

\subsubsection{Broad Band Spectrum}

\begin{figure}
\epsfig{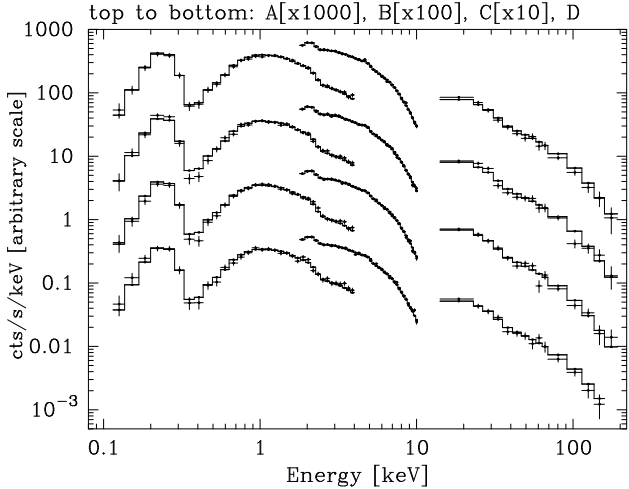}
\caption {Data for the four observations. Observations A to D 
from top to bottom, respectively.  For clarity, datasets A, B and 
C are scaled in count rate by 3, 2 and 1 orders of magnitude.
\label{fig:ldata} }
\end{figure}

We jointly fit the data
of the three instruments, for each of the four observations. The model
adopted is a single power law with Galactic absorption
(N$_{\rm H} =1.69\times 10^{20}$ cm$^{-2}$, Lockman \& Savage 1995). 
We add two further parameters, which we allow to be free, accounting for
the mis-calibration of the LECS and PDS with respect to the MECS, which
turned out to be the best calibrated instrument following independent tests.

\begin{figure}
\epsfig{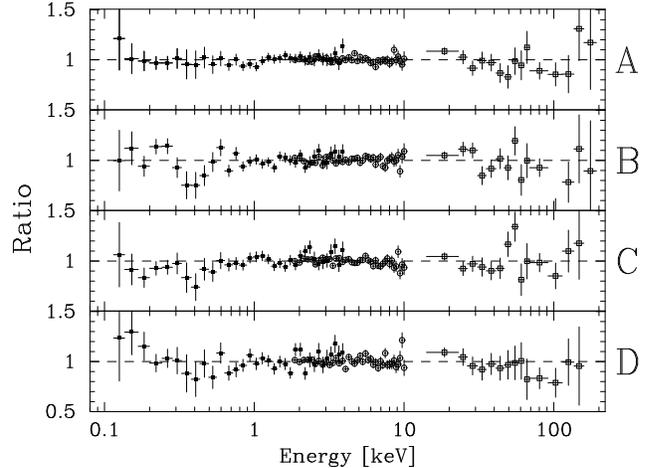}
\caption {Data/model ratio for the four observations. 
The fit model is a power law with Galactic absorption, with
free normalization among the different instruments. 
\label{fig:ratio-all} }
\end{figure}

The data and the data/model ratio are plotted in Fig.~\ref{fig:ldata} 
and Fig.~\ref{fig:ratio-all}, respectively, while the best fit
parameters are listed in Table~\ref{tab:broad-band-fits}.  
As the fit to the broad band spectrum is dominated, in terms
of $\chi^2$ statistics, by the MECS data points, the best fit value of
the spectral index is essentially that of the MECS data 
(see next \S3.2.3). 
As shown in Table~\ref{tab:broad-band-fits}, a single absorbed power law
is an acceptable representation of the data for observations A and C, it
is marginally consistent with data set D, while it is rejected at 99.9\%
level in observation B. 
Visual inspection of Fig.~\ref{fig:ratio-all} shows that large scatter
is present in the PDS data, which also appear to be systematically steeper
than the MECS data. 
A feature around $\sim 0.4$ keV is present in the LECS data. 

It is worth noting that, though the relative LECS/MECS and PDS/MECS
normalizations are marginally consistent with the fiducial ranges as given by
the SDC (0.7--0.75 and 0.8--0.85, respectively), the LECS/MECS ratio possibly varies  
among different datasets. One should be careful in interpreting variations 
of the LECS/MECS normalization ratio, as the presence of the strongback and of 
a support grid structure in the LECS window can actually affect such quantity 
(Parmar et al. 1997). Recently, an in--flight raster scan of the central 
$(7'\times 7')$ area of the LECS and MECS
detectors (14 short observations of LMCX1) has allowed to map 
the LECS/MECS count ratio in the 2--8 keV band as a function of the position of
the source in the LECS field of view. The scan has shown that the LECS/MECS 
ratio can 
decrease by about 22$\%$
across the detector (Ricci 1998, private comunication).
However, such effect occurs when the centroid of the source lies at least at 
$1.2'$
from the LECS center.
Since 3C 273 was always in the center of the field of view during the AO1
observations, and the positions of the source centroid do not differ by more 
than
$20''$, it is unlikely that window support obscuration affects
the LECS/MECS ratio in our datasets. 

Nevertheless we must stress that fiducial values apply only under the
assumption that the fitted model is a good description of the data
throughout all the three instruments.
If the model fails to describe the data in one of the instruments, this
affects the resulting value of the inter--calibration parameter.

We therefore conclude that a single power law, which is essentially
determined by the MECS statistics, is not a good description of the data
in the LECS and PDS instrument ranges.

		%---------------------------%

\subsubsection{LECS}

%---------------------TABLE 3---------------------------------------------
\begin{table*}
\caption[]{Power Law Fits: LECS\label{tab:fit-lecs}}
\begin{tabular}{lcccc}
\noalign{\hrule}
\multicolumn{1}{l}{Obs.} &           
\multicolumn{1}{c}{N$_{\rm H}$}&
\multicolumn{1}{c}{$\Gamma$} &
\multicolumn{1}{c}{F$_{[0.2-1]\rm{keV}}$}&            
\multicolumn{1}{c}{$\chi^{2}/d.o.f$ ($\chi^{2}_\nu$)}\\
&&&&\\
&\multicolumn{1}{c}{($10^{20}$ cm$^{-2}$)}
&&
\multicolumn{1}{c}{($10^{-10}$ erg/cm$^2$/s)}&\\
\noalign{\hrule}
&&&&\\
{A} &  { 1.69 (f)} & { $1.54_{-0.03}^{+0.02}$ } & 
{ $0.337$ } & { 12.6/32 (0.40) } \\
{ } &  { $1.55_{-0.26}^{+0.27}$ } & { $1.52_{-0.05}^{+0.05}$ } & 
{ $0.336$ } & { 11.3/31 (0.37) } \\
&&&&\\ 
{B} &  { 1.69 (f)} & { $1.57_{-0.03}^{+0.02}$ } &
{ $0.341$ } & { 42.2/32 (1.32) } \\
{ } &  { $1.35_{-0.25}^{+0.27}$} & { $1.58_{-0.05}^{+0.05}$ } &
{ $0.340$ } & { 35.4/31 (1.14) } \\
&&&&\\
{C} &  { 1.69 (f)} & { $1.56_{-0.03}^{+0.03}$ } &
{ $0.332$ } & { 25.1/32 (0.78) } \\
{ } &  { $1.70_{-0.30}^{+0.31}$} & { $1.56_{-0.05}^{+0.05}$ } &
{ $0.226$ } & { 25.1/31 (0.81) } \\
&&&&\\
{D} &  { 1.69 (f)} & { $1.54_{-0.04}^{+0.03}$ } &
{ $0.219$ } & { 37.4/32 (1.17) } \\
{ } &  { $1.24_{-0.33}^{+0.36}$} & { $1.48_{-0.06}^{+0.06}$ } &
{ $0.215$ } & { 30.7/31 (0.99) } \\
&&&&\\
\noalign{\hrule}
\end{tabular}
{\raggedright\par
{\bf Note:} quoted errors are 90\% confidence 
intervals for one ($\Delta\chi^2=2.706$) or two interesting parameters
($\Delta\chi^2=4.61$). \par }
\end{table*}
%------------------------------------------------------------------------

The LECS spectra for the four observations were fit by a simple power
law with free absorption. Data/model ratios are plotted in Fig.~6,
while the best fit parameters are presented in Table~\ref{tab:fit-lecs}. 
All the four data sets can be fit reasonably well by the
absorbed power--law model.  The 90\% confidence regions for N$_{\rm H}$ and
$\Gamma$ are shown in Fig.~\ref{fig:lecs_cont}.  
The spectral index is also plotted in Fig.~\ref{fig:indices}, for
N$_{\rm H}$ =  N$_{\rm H,gal}$. 
The value of $\Gamma$ is in general smaller than that  
obtained in the broad band fit with fixed Galactic absorption. 
Observations B and D require a value of N$_{\rm H}$ 
slightly lower than the Galactic absorption column density. 
This could indicate that a weak
soft excess is present in the data, although the evidence is, at best, 
marginal. In fact the spectral indices for
all the four observations are, within the errors, consistent with the
average value $\Gamma \simeq 1.53$ (see also Fig.~8). No apparent
correlation of N$_{\rm H}$ with the flux is found.

\begin{figure}
\epsfig{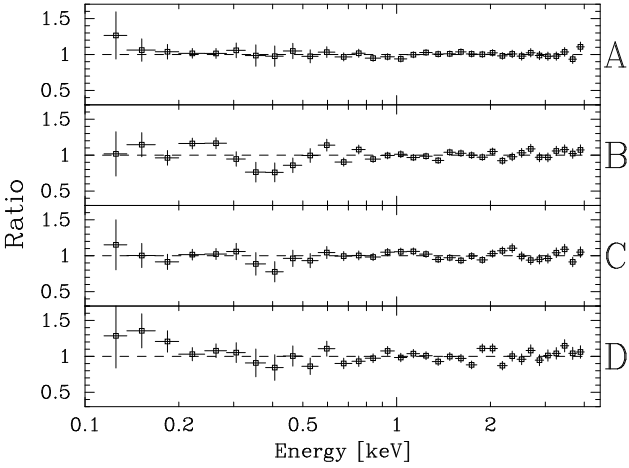}
\caption {Data/model ratio of power law fits to LECS data.
\label{fig:ratio-lecs} }
\end{figure}

As evident from Fig.~5 and 6, in all the observations other
than A, most of the contribution to the $\chi^2$ is due to residual
structures at $\sim 0.3-0.5$ keV.  It is known that they are spurious
and produced by a mis--calibration of the LECS in the Carbon edge
region (Fossati \& Haardt 1997, Guainazzi \& Grandi 1997, Orr et
al. 1997).  The latest LECS response matrices coupled with the new
data reduction pipeline strongly reduce the amplitude of
these features, but do not completely remove them. Some excess of
residuals is also present near $\sim 2.2$ keV, probably due to
inaccuracies in the calibration of the instrumental gold feature.

\begin{figure}
\epsfig{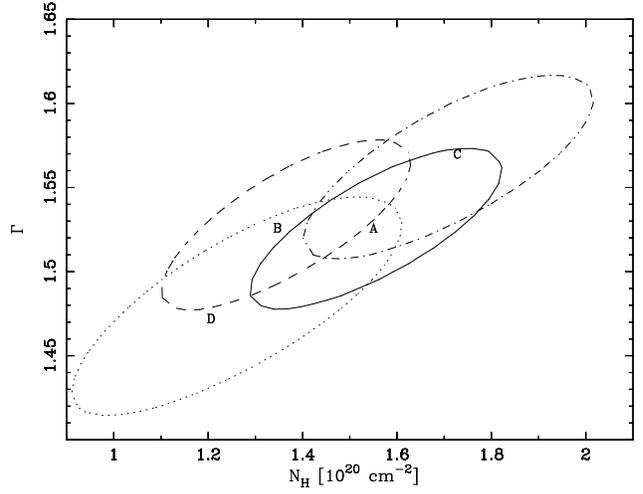}
\caption {90\% contour regions of column density and spectral index
for absorbed power law fits to LECS data. Solid, dashed, dot--dashed,
and dotted lines represent observations A, B, C, and D, respectively. 
The position of the labels within each region marks the best fit 
values. \label{fig:lecs_cont} }
\end{figure}

Therefore, within these uncertainties, LECS data are consistent with a
power--law spectrum absorbed by Galactic material along the line of
sight. In two out of four observations (B and D) there is some marginal 
indication of the presence of a soft excess. There is no evidence of
changes of the spectral index among the four observations.

\begin{figure}
\epsfig{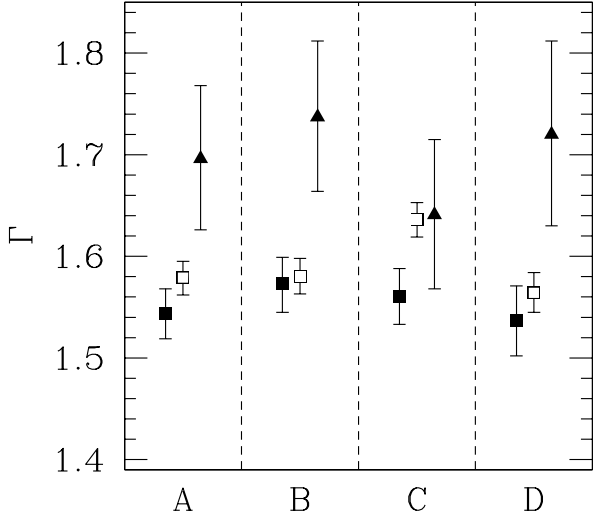}
\vskip -1.1 cm
\caption {LECS (filled squares), MECS (open squares) and PDS (filled
triangles) spectral indices for the four observations. 
Error bars are 90\% intervals for one interesting parameter.
\label{fig:indices} }
\end{figure}

\subsubsection{MECS}

%---------------------TABLE 4---------------------------------------------
\begin{table}
\caption[]{Power Law$^{\rm a}$ Fits: MECS\label{tab:fit-mecs}}
\begin{tabular}{lccc}
\noalign{\hrule}
\multicolumn{1}{l}{Obs.} &           
\multicolumn{1}{c}{$\Gamma$} &
\multicolumn{1}{c}{F$_{[2-10]\rm{keV}}$}&            
\multicolumn{1}{c}{$\chi^{2}/d.o.f$ ($\chi^{2}_\nu$) }\\
&&&\\
&&\multicolumn{1}{c}{($10^{-10}$ erg/cm$^2$/s)}&\\
\noalign{\hrule}
&&&\\
{ A } & { $1.58_{-0.02}^{+0.02}$ } & { $1.18$ } & { 29.0/37 (0.78) } \\ 
&&&\\ 
{ B } & { $1.58_{-0.02}^{+0.02}$ } & { $1.14$ } & { 42.1/37 (1.14) } \\
&&&\\ 
{ C } & { $1.64_{-0.02}^{+0.02}$ } & { $1.07$ } & { 36.7/37 (0.99) } \\ 
&&&\\ 
{ D } & { $1.56_{-0.02}^{+0.02}$ } & { $1.03$ } & { 48.7/37 (1.32) } \\
&&&\\
\noalign{\hrule}
\end{tabular}
{\raggedright\par
{\bf Note:} quoted errors are 90\% confidence 
intervals for one interesting parameter ($\Delta\chi^2 = 2.706$).\par\noindent 
$^{\rm (a)}$ N$_{\rm H}$ fixed at LECS best fit value
(see Table~\ref{tab:fit-lecs}). \par }
\end{table}
%--------------------------------------------------------------------------

MECS data of all of the four observations are well described by a single
power law. Data/model ratios and best fit spectral indices are shown
in Fig.~9 and 8 respectively, and their values are reported in
Table~\ref{tab:fit-mecs}. Cold gas absorption column density was fixed at the
appropriate LECS value in each data set. In any case, its precise
value does not affect the fit of the MECS data, as N$_{\rm H}$ is always
lower than several times $10^{20}$ cm$^{-2}$.

With an average flux of $\simeq 1.14\times 10^{-10}$ erg/cm$^2$/s [2--10
keV], the source was roughly in the middle of the range of fluxes
historically observed ($\sim 0.6-1.7 \times 10^{-10}$ erg/cm$^2$/s) and the
corresponding unabsorbed 2--10 keV luminosity in the Quasar frame is
$\sim 1.3 \times 10^{46} h_{50}^{-2}$ erg/s ($h_{50}=H_0$/50 km/s/Mpc, 
$q_0=0$).

\begin{figure}
\epsfig{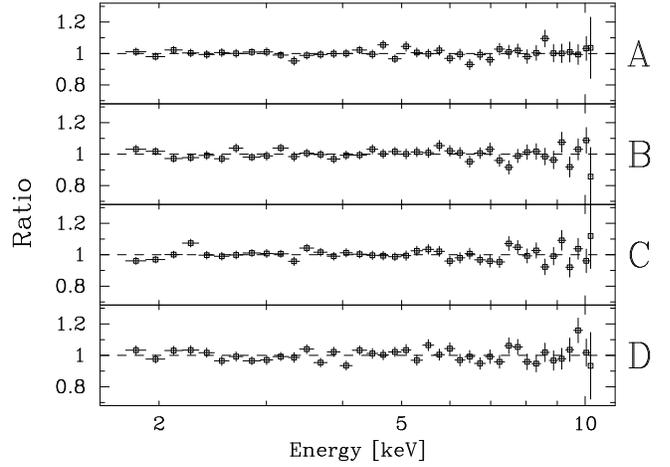}
\caption {Data/model ratio of power law fits to MECS data.
\label{fig:ratio-mecs} }
\end{figure}

The energy spectral index is rather flat, with $\Gamma\simeq 1.55$ for
three of the four observations (see Table~\ref{tab:fit-mecs}), consistent
with the {\it GINGA} results (Lawson \& Turner 1996), but slightly flatter
than {\it ASCA} 1994 observations (Yaqoob et al. 1994). The spectrum during
observation C is instead steeper ($\Gamma=1.62\pm{0.02}$). Given the
small errors in the determination of the slope, the variation of
spectral index, albeit small, is statistically significant. It is also
important to note that the spectral index in the MECS range is only
barely consistent with the index derived for the lower LECS energy
range (see Fig.~8).

\begin{figure}
\epsfig{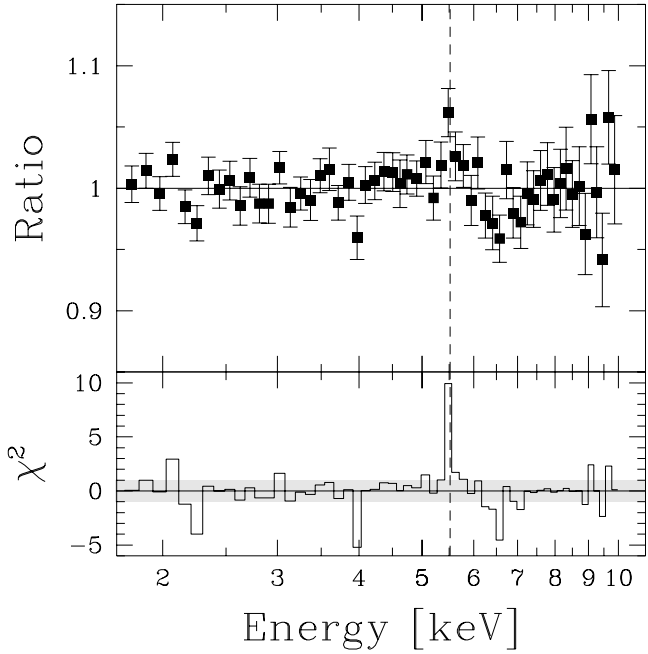}
\caption {Data/model ratio (upper panel) and contribution to $\chi^2$
(lower panel) when the co--added MECS ratios (the data) are fit with a
constant=1 (the model). The dashed vertical line is at the energy of a
6.4 keV (source frame) iron emission line. 
\label{fig:iron-line} }
\end{figure}

The inclusion of an iron emission line (at 6.4 keV) does not statistically 
improve the fit in any of the four observations. 
All the available MECS counts could be added in a single
spectrum to improve the statistics but, in view of the difference
existing in the value of $\Gamma$ between observation C and the other
three, we chose to sum the data normalized to the best fit of each
observation, with the errors propagated accordingly.  
Furthermore, as weak narrow features could be smoothed out using a bin
size that is too wide, in order to test for emission lines we binned the
spectrum using the {\tt grouping3} template, which allows for a larger
number of bins than {\tt grouping2}.
As shown in Fig.~10, both the shape of the
residuals with respect to a constant ($=1$), and the sum of the ratios
itself, clearly indicate the presence of an emission line at the source 
frame energy of 6.4 keV.  
The significance of such a feature is above $3\sigma$, and its equivalent
width is estimated to be $\simeq 20$ eV.  
Although the residuals in Fig.~\ref{fig:iron-line} might be even suggestive
of the presence of an iron absorption edge at the expected energy, the
statistics is not good enough to discriminate this feature from the
noise.

\subsubsection{PDS}

%---------------------TABLE 5---------------------------------------------
\begin{table}
\caption[]{Power Law Fits: PDS\label{tab:fit-pds}}
\begin{tabular}{cccc}
\noalign{\hrule}
\multicolumn{1}{l}{Obs.} &           
\multicolumn{1}{c}{$\Gamma^{\rm a}$} &
\multicolumn{1}{c}{F$_{[20-100]\rm{keV}}$}&            
\multicolumn{1}{c}{$\chi^{2}/d.o.f$ ($\chi^{2}_\nu$)  }\\
&&&\\
&&\multicolumn{1}{c}{($10^{-10}$ erg/cm$^2$/s)}&\\
\noalign{\hrule}
&&&\\
{A} & { $1.70_{-0.07}^{+0.07}$ } & { $2.58$ } & { \phantom{8}8.9/13 (0.69) } \\ 
&&&\\ 
{B} & { $1.74_{-0.07}^{+0.08}$ } & { $2.62$ } & { 20.1/13 (1.55) } \\ 
&&&\\ 
{C} & { $1.64_{-0.07}^{+0.07}$ } & { $2.32$ } & { 16.3/13 (1.25) } \\ 
&&&\\
{D} & { $1.72_{-0.09}^{+0.09}$ } & { $2.17$ } & { \phantom{2}2.6/12 (0.22) } \\ 
&&&\\
\noalign{\hrule}
%% \multicolumn{4}{l}{\bf Note: quoted errors are 90\% confidence for 
%% one interesting parameters ($\Delta\chi^2=2.706$). }
\end{tabular}
{\raggedright\par
{\bf Note:} quoted errors are 90\% confidence 
intervals for one interesting parameters ($\Delta\chi^2 = 2.706$).\par }
\end{table}
%------------------------------------------------------------------------

\begin{figure}
\epsfig{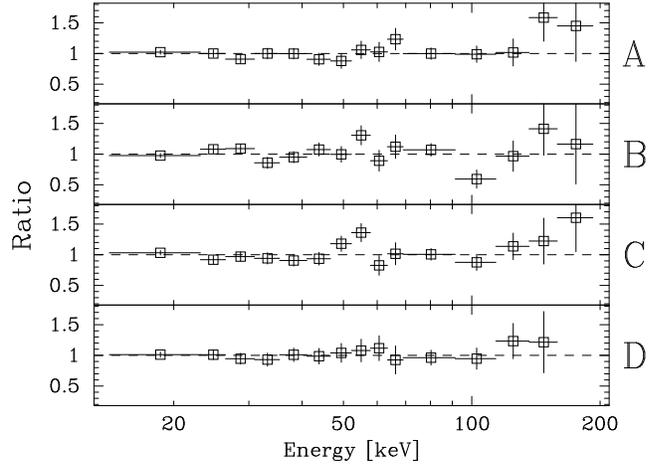}
\caption {Data/model ratio of power law fit to PDS data.
\label{fig:ratio-pds} }
\end{figure}

Power law fits to the four PDS datasets alone are satisfactory in observations 
A, C and D (see Fig.~\ref{fig:ratio-pds} and Table~\ref{tab:fit-pds}), 
despite a large
scatter in the data points (probably related to a not--well calibrated
background subtraction). The exceedingly large $\chi^2$ value of 
the fit to observation B is essentially due to a single bin at 
$\simeq 100$ keV, and, again, it is probably background related. 
The large error in the determination of the slope does not allow to
detect spectral variations in the PDS data. 
The PDS spectral indices we derived are consistent with the OSSE results 
(McNaron--Brown et al. 1995). However, as shown in
Fig.~8, they are statistically steeper 
than the MECS values in at least three out of four observations. 

We also analyzed the PDS spectra extracted using the {\tt SAXDAS} 
software package, which, among other differences, 
does not includes the rise--time correction 
(see Section 2.4). While the spectral
indices are consistent with those reported in Table~\ref{tab:fit-pds}, the
90\% confidence range is systematically shifted to lower values for all the
four observations, and marginally consistent with the MECS slopes.

In conclusion, the steepening of the broad band spectrum
towards higher energies (Fig.~\ref{fig:indices}) is evident using the
pipelines of both of the two data 
reduction softwares, though the significance and degree of such a trend 
depends somewhat on the adopted PDS data reduction technique. 

		%-----------------------%

\subsection{SVP spectral analysis}

It is useful to compare the results of the \BS AO1 observations with the
previous \BS SVP data, presented in Grandi et al. (1997) where the interested
reader may find a detailed description of the observations and data reduction. 

We have repeated the LECS and MECS analysis of the SVP data taking
advantage of the revised response matrices. 
The Grandi et al. results are, within the errors, quantitatively confirmed
(see Table~\ref{tab:fit-svp}).  

The MECS continuum is described by a power law which is marginally steeper 
than that observed in Jan. 1997, while the flux is a 
factor $\sim 1.6$ lower. 
MECS data also reveal a weak emission line interpreted as fluorescence
K$_{\alpha}$ transition of cold or mildly ionized iron. 
In order to estimate the line equivalent width, we operated as follows:
first, we obtained the slope of the continuum by fitting the data excluding 
the 5--6 keV interval (model (e) in Table~\ref{tab:fit-svp}), 
where the redshifted 6.4 keV line shows up; then we fit the continuum
plus emission line keeping frozen the power law index, and the line energy
and intrinsic width (model (f) in Table~\ref{tab:fit-svp}).

\begin{figure}
\epsfig{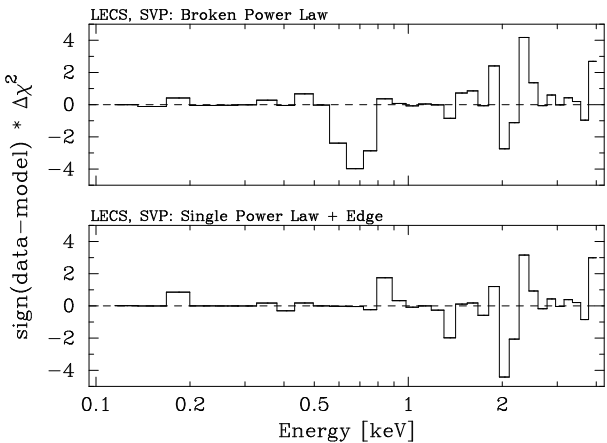}
\caption{Contributions to $\chi^2$ for fits to SVP LECS data:
broken power law (top panel), and single power plus absorption edge 
(bottom panel). 
\label{fig:svp-chisq} }
\end{figure}

In the LECS data a clear absorption feature and possibly a soft excess below
0.5 keV, are superimposed to the power law.
The fit to the 0.2--4 keV data with a broken power
law (the slope at energies $\gta 0.55$ keV is kept fixed at the MECS value) 
is not statistically distinguishable from a fit with a single steep
power law plus absorption edge (see Table~\ref{tab:fit-svp}).
However, as shown in Fig.~\ref{fig:svp-chisq} the larger value of $\chi^2$
in the broken power law fit is entirely due to extra contributions
occurring in the 0.5--0.7 keV range, i.e. there is a systematic 
defect at the energy of the absorption feature.

\begin{figure}
\epsfig{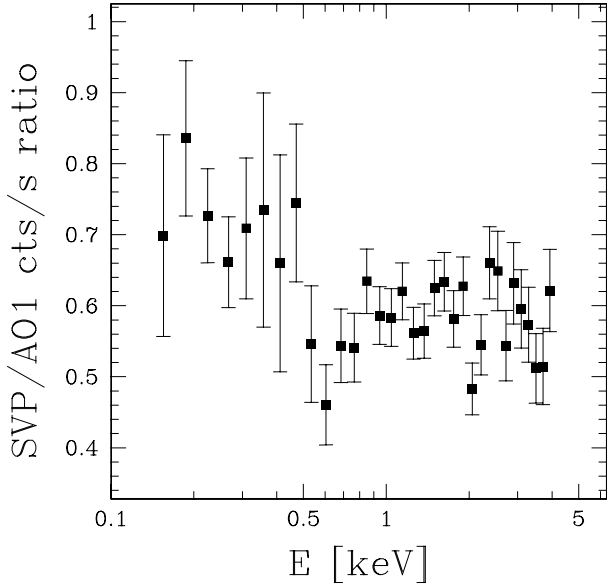}
\caption {Ratio of SVP and AO1 LECS data. Data/model ratios 
from observations A, B, C, and D of AO1 are added together.
\label{fig:ratio-svp-ao1} }
\end{figure}

To further check the
reliability of the absorption feature and soft excess detection, we
computed the ratio between the SVP and the AO1 LECS data, as the ratio
is certainly less subjected to instrumental effects.  
Fig.~\ref{fig:ratio-svp-ao1} unambiguously indicates that the soft
excess and the absorption edge at $\simeq$ 0.6 keV are genuinely present
in the SVP data only.

%---------------------TABLE 6---------------------------------------------
\begin{table*}
\caption[]{SVP Data Spectral Fits\label{tab:fit-svp}}
\begin{tabular}{llcccccc}
\noalign{\hrule}
\multicolumn{1}{l}{LECS} &
\multicolumn{1}{c}{Model} &
\multicolumn{1}{c}{N$_{\rm H}$}&
\multicolumn{1}{c}{$\Gamma$} &
\multicolumn{1}{c}{E$_{\rm break}^{\rm g}$/E$_{\rm edge}^{\rm g}$} &
\multicolumn{1}{c}{$\Gamma_{\rm high}/\tau_{\rm edge}$} &
\multicolumn{1}{c}{F$_{[0.2-1]\rm{keV}}$}&            
\multicolumn{1}{c}{$\chi^{2}/d.o.f$ ($\chi^{2}_\nu$)}\\
{} & {} &\multicolumn{1}{c}{($10^{20}$ cm$^{-2}$)} & & { (keV) }& &
\multicolumn{1}{c}{($10^{-10}$ erg/cm$^2$/s)}&\\
\noalign{\hrule}
&&&&&&&\\
% \multicolumn{7}{l}{LECS} \\
% &&&&&&&\\
  % \multicolumn{7}{l}{single power law} \\
{}&{PL$^{\rm a}$} &  { 1.69 (f)} & { $1.623_{-0.039}^{+0.036}$ } & { ....
} & { .... } & { 0.152 } & { 43.2/32 (1.35) }\\
&&&&&&&\\ 
{}&{PL+N$_{\rm H}$$^{\rm b}$} &  { $1.07_{-0.31}^{+0.35}$ } & 
{ $1.537_{-0.068}^{+0.068}$ } & { .... } & { .... } & { 0.148 } & { 29.4/31
(0.95) }\\
&&&&&&&\\ 
{}&{BKN PL$^{\rm c}$} &  { 1.69 (f)} & { $1.879_{-0.143}^{+1.038}$ } & 
{ $0.61_{-0.43}^{+0.16}$ } & { 1.599 (f) } & 
{ 0.154 } & { 30.7/31 (0.99) }\\
&&&&&&&\\ 
{}&{PL+edge$^{\rm d}$} &  { 1.69 (f)} & { $1.671_{-0.060}^{+0.055}$ } & 
{ $0.64_{-0.10}^{+0.14}$ } & { $0.57_{-0.36}^{+0.43}$ } & 
{ 0.149 } & { 24.0/30 (0.80) }\\
&&&&&&&\\
\noalign{\hrule}
{MECS} &
{Model} &
{N$_{\rm H}$}&
{$\Gamma$} &
{E$_{\rm line}^{\rm g}$} &
{EW$_{\rm line}^{\rm g}$} &
{F$_{[2-10]\rm{keV}}$}&            
{$\chi^{2}/d.o.f$ ($\chi^{2}_\nu$)}\\
{} & {} & { ($10^{20}$ cm$^{-2}$) } & & { (keV) }& { (eV) } &
{ ($10^{-10}$ erg/cm$^2$/s) } & {} \\
\noalign{\hrule}
&&&&&&&\\
 % \multicolumn{7}{l}{MECS} \\
 % &&&&&&&\\ 
{}&{PL$^{\rm e}$ } &  { 1.69 (f)} & { $1.599_{-0.011}^{+0.010}$ } & { .... } & 
{ .... } & { 0.670 } & { 41.1/32 (1.28) }\\
&&&&&&&\\ 
 % \multicolumn{7}{l}{single power law + emission line} \\
{}&{PL+line$^{\rm f}$} &  { 1.69 (f)} & { 1.599 (f) } & 
{ 6.4 (f) } & { $29.2_{-11.6}^{+5.7}$ } & 
{ 0.673 } & { 46.9/37 (1.27) }\\
&&&&&&&\\ 
\noalign{\hrule}
\end{tabular}
{\raggedright\par
{\bf Note:} quoted errors are 90\% confidence intervals for one
($\Delta\chi^2=2.706$), two ($\Delta\chi^2=4.61$),
or three interesting parameters ($\Delta\chi^2=6.25$). \par\noindent
$^{\rm (a)}$ single power law with Galactic absorption.\par\noindent
$^{\rm (b)}$ single power law with free absorption.\par\noindent
$^{\rm (c)}$ broken power law with Galactic absorption.\par\noindent
$^{\rm (d)}$ single power law with Galactic absorption plus absorption
edge.\par\noindent
$^{\rm (e)}$ single power law with Galactic absorption, excluding the
data in the range 5--6 keV.\par\noindent
$^{\rm (f)}$ single power law with Galactic absorption plus emission
line. The line width is fixed at 0.1 keV, .\par\noindent
$^{\rm (g)}$ Energy in the source frame.\par\noindent
}
\end{table*}
%------------------------------------------------------------------------

Note also that the feature detected in the AO1 LECS data at 0.4 keV
vanishes almost completely in the SVP/AO1 ratio. This implies that
either it is equally present in both datasets or has an instrumental
origin. We performed a survey in a sample of LECS datasets of
different blazars (Fossati \& Haardt 1997) and concluded that the
latter option is the most likely.

% The "genuine" SVP absorption feature can be modeled with an edge,
% yielding an (source frame) energy E=0.64 keV and an optical depth
% $\tau$=0.57. 
% The effect of the new LECS matrix is to
% shift the edge to higher energies and reduce its depth. However the
% associated uncertainties are so large that the new parameters are
% still consistent with those reported by Grandi et al. (1997). A
% slightly better fit is obtained by modeling the feature with an
% Gaussian absorption line (at $E=0.64_{-0.07}^{+0.09}$ keV) with EW
% $\simeq 48_{-33}^{+37}$ eV.

		%-----------------------%

\section{Simultaneous observations\label{sec:simult}}

3C 273 has been constantly monitored by BATSE during the last few years, and it
is of some interest to frame the \BS SVP and AO1 observations into such a long 
look of the source. 

\begin{figure}
\epsfig{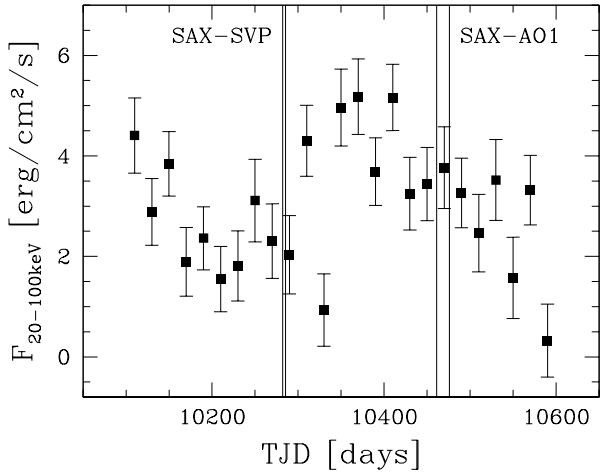}
\vskip -2.5 cm
\caption {BATSE light curve. Twenty days intervals resulting from
the rebinning of the daily data obtained with LADs in Earth Occultation Mode.
The two vertical strips represent the windows of \BS observations during the 
SVP and the AO1. \label{fig:batse_lc} }
\end{figure}

In Fig.~\ref{fig:batse_lc} we show the BATSE light curve over 500 days
covering both of the \BS observations. The fluxes in the 20--100
keV energy band have been obtained by folding a single power law of photon
index 1.7, after the PDS results. The BATSE data were collected by the
Large Area Detectors (LADs) in Earth occultation mode. The occultation of a
hard X-ray source in the sky by the Earth produces a step like feature
superposed to the continuous detector background count rate. A pair of
rising and setting steps are generated, and the standard Earth occultation
technique (Harmon et al. 1992) has been applied to monitor X-ray sources on
a daily basis. In the case of 3C~273 a bin of 20 days of the daily data has
been applied in order to improve the statistics and enlight the trend
of the source on longer time scales. 

Despite the fact that local minima or maxima are present in the BATSE light
curve on monthly scale, Fig.~\ref{fig:batse_lc} shows that the \BS SVP
observation was performed during a period of long term increase of the
source, while the AO1 observations occurred in a period of long term decrease. 

We also performed an optical monitoring during the AO1 observations. The
optical data were taken from Jan. 12 to Jan. 31 at the Torino
Astronomical Observatory, with the $1.05 \, \rm m$ telescope REOSC. The
instrumentation comprises a $1242 \times 1152$ CCD camera with $0.47 \, \rm
arcsec$ per pixel scale and standard Johnson-Cousins $BVR$ filters. Typical
exposure times for 3C 273 are 180 s in the R band, 240 s in V, and 300 s in B.
Data were analyzed with the Robin procedure locally developed, which performs
bias subtraction, flat field correction and circular Gaussian fit after
background subtraction. The source magnitude is obtained by relative photometry
with respect to the reference stars calibrated by Smith et al. (1985). Errors
are of the order of a few hundredths of magnitude. Optical fluxes were
calculated using the absolute flux calibration by Bessell (1979); no correction
for galactic extinction is needed for 3C 273. 

The R, B and V average magnitudes are 12.7, 13.0 and 13.1, respectively,
corresponding to an integrated flux in R and V of $\simeq 1.2\times 10^{-10}$
erg/cm$^2$/s, of $\simeq 1.6\times 10^{-10}$ erg/cm$^2$/s in B. 
These values are
slightly larger than the flux in the 2--10 keV band (see Table 4). It worths
noting that no optical variability at a level larger than a few percent has
been detected during the monitoring. 

		%-----------------------%

\section{Discussion: The hidden Seyfert nucleus\label{sec:discussion}}

As reported in the previous section, the AO1 data are only marginally
consistent with a simple single power law model with Galactic absorption. There
are (admittedly weak) indications of features, namely iron emission line and a
high energy curvature of the \BS spectra.  More importantly, the presence of
these same (but stronger) features, absorption around 0.6 keV, 
and a soft excess component at very soft energies during the weakest SVP state,
lead us to consider a more complex underlying (variable) emission model. 

In particular, it has been already (qualitatively) suggested (see
references in the Introduction, Cappi et al. 1998) that an iron
emission line in the spectra of radio--loud quasars might indicate
that the X--ray emission comprises two components: a Seyfert--like
thermal X--ray spectrum, characterized by features induced by
reflection over optically thick cold material subtending $\sim 2\pi$
to the `primary' source, and a flat non--thermal component, arising as
emission from plasma moving at relativistic speed in the jet, as
typical of radio--loud objects.

We therefore modelled the data with a complex model, consisting in the 
sum of a constant `thermal' component and a variable non--thermal power law. 
Clearly our aim is {\it not} to obtain a
significantly better fit for the AO1 data, but rather to determine
whether the variations in the observed features between the different
observations are consistent with the scenario of an accreting object
which dissipates energy in an accretion disc and its corona but where
also a powerful jet is formed producing extremely variable beamed
emission.

The thermal spectrum has been represented by parameters typical of
Seyfert spectra (e.g. Nandra \& Pounds 1994) with a primary $\Gamma=1.9$
power--law emission impinging on cold matter subtending a 2$\pi$ solid 
angle, giving rise to an iron line at 6.4 keV (source frame) with 
an EW of 150 eV. In order to establish the absolute normalization of the 
thermal component, we considered the SVP data.
The flux of the underlying Seyfert--like spectrum is determined by the
simple requirement that the intrinsic 150 eV of the iron line EW are observed,
once diluted by the non--thermal component, as $\sim 30$ eV.
This sets the flux of the Seyfert--like component at 6.4 keV (source frame) 
in the SVP to be 1/5 of the SVP total observed flux at this energy.
We then assume that such flux level characterizes the Seyfert--like component
in all the source states, i.e. the Seyfert--like component did not change
among SVP and AO1 data.
In other words, we test the hypothesis that all of the spectral differences
observed among SVP and AO1 observations are accounted for by the variable
non--thermal component. The latter has been modelled as a power--law with
a (parabolic) turnover, to mimic the cutoff around a few MeV, detected 
in the $\gamma$--ray data (e.g. McNaron--Brown et al. 1995). 

\begin{figure}
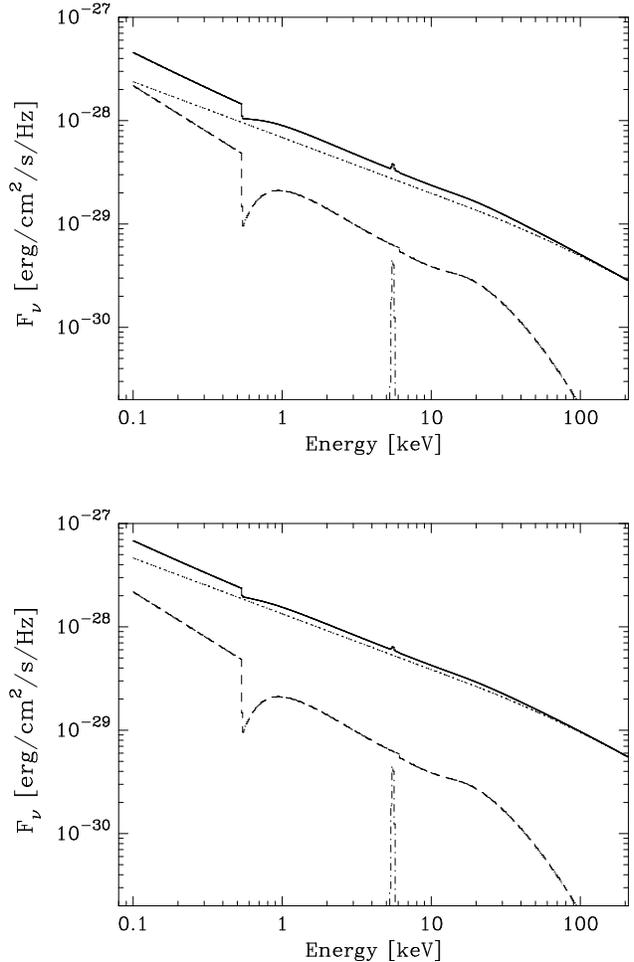

% \hbox{
\epsfig{figure=7883.f15A,width=6.7cm,angle=-90}   % \hfill
\epsfig{figure=7883.f15B,width=6.7cm,angle=-90} 
% }
\caption {Best fit model for the SVP and AO1--A data 
(top and bottom panel, respectively). 
The dashed and dash--dotted lines indicate the
thermal reflected spectrum and the iron emission line, the dash--dotted one
represents the non--thermal power--law emission and the solid line is
their sum.
\label{fig:models} }
\end{figure}

\begin{figure}
\epsfig{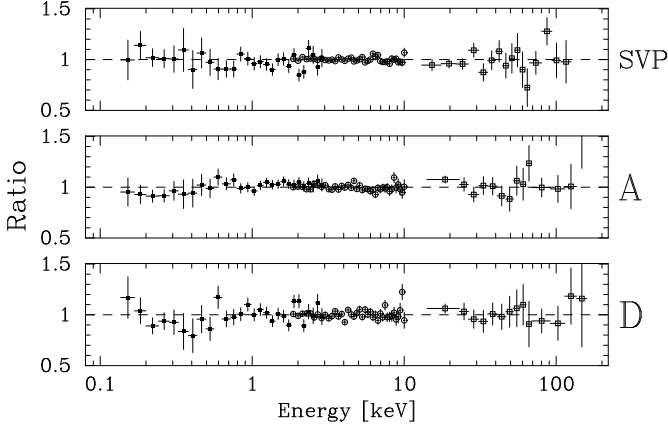}
\caption {Data to model ratios for the SVP and AO1 A and D datasets for the 
best fit parameters reported in Table~\protect\ref{tab:fit_modellone}.
\label{fig:3datasets} }
\end{figure}

%---------------------------TABLE 7----------------------------------------
\begin{table}
\caption[]{Fit with Seyfert+Jet Model$^{\rm a}$\label{tab:fit_modellone} }
\begin{tabular}{@{}lcc}
\noalign{\hrule}
{ Parameter } & { Value } & { Units } \\
\noalign{\hrule}
\multicolumn{3}{c}{\underbar{General Parameters}}\\
&&\\
{ \# N$_{\rm H}$ }          & { 1.69 }  & { (10$^{20}$ cm$^{-2}$) } \\
{ \# z }                    & { 0.158 } & \\
{ \phd LECS/MECS$^{\rm b}$ }& { $0.75^{+0.01}_{-0.01}$ }&          \\ 
{ \phd PDS/MECS$^{\rm b}$ } & { $0.85^{+0.03}_{-0.03}$ }&          \\ 
\noalign{\hrule}
\multicolumn{3}{c}{\underbar{Seyfert Parameters}}\\
&&\\
{ \# R }                  & { 1 }     & {  } \\
{ \# E$_{\rm fold}$ }     & { 100 }   & { (keV) } \\
{ \# cos(i) }             & { 0.95 }  & \\
{ \# $\Gamma$ }           & { 1.9 }   & \\
{ \# E$_{\rm line}$ }     & { 6.4 }   & { (keV) } \\
{ \# $\sigma_{\rm line}$ }& { 0.1 }   & { (keV) } \\
{ \# F$_{\rm line}^{\rm c}$ }& { $3 \times 10^{-5}$ }&{ (photons/cm$^2$/s) }\\
{ \phd E$_{\rm edge}^{\rm b}$ }   &{ $0.62^{+0.12}_{-0.12}$ }&{ (keV) } \\ 
{ \phd $\tau_{\rm edge}^{\rm b}$ }&  { $1.7^{+0.6}_{-0.7}$ }   &          \\ 
{ \# F$_{[\rm 2-10keV]}$ }   & { 0.129 } & { (10$^{-10}$ erg/cm$^2$/s) } \\
\noalign{\hrule}
\multicolumn{3}{c}{\underbar{Jet Parameters}}\\
&&\\
{ \phd $\Gamma^{\rm b}$ }        & { $1.54^{+0.01}_{-0.01}$ } &           \\ 
{ \phd E$_{\rm break}^{\rm b}$ } & { $16.6^{+21.4}_{-8.0}$ } & { (keV) } \\ 
{ \# E$_{\rm peak}^{\rm a}$ }    & { 2.0 }                    & { (MeV) } \\ 
{ \phd F$_{[\rm 2-10keV]}$ }     &           &              \\
{ \phantom{ABCD}SVP: }  & { 0.544 } & { (10$^{-10}$ erg/cm$^2$/s) } \\
{ \phantom{SVPDAD}A: }  & { 1.060 } & { (10$^{-10}$ erg/cm$^2$/s) } \\
{ \phantom{SVPDAD}D: }  & { 0.892 } & { (10$^{-10}$ erg/cm$^2$/s) } \\
\noalign{\hrule}
{ $\chi^2/d.o.f.$ ($\chi^2_\nu$) } & \multicolumn{2}{c}{ 230.6/238 (0.969) }\\ 
\noalign{\hrule}
\end{tabular}
{\raggedright\par
{\bf Note:}  \par\noindent
$^{\rm (a)}$ Parameters marked with a '\#' have not been allowed to vary
during the fit.
The reported best--fit values and errors are those obtained keeping the
value of the ``Peak Energy" (E$_{\rm peak}$) fixed at 2 MeV. 
The only parameter significantly affected by a change in E$_{\rm peak}$
is E$_{\rm break}$.  
However for E$_{\rm peak}$ moving in the range 1 -- 100 MeV it always
holds $5 \le$ E$_{\rm break} \le 30-40$ keV. 
\par\noindent 
$^{\rm (b)}$ quoted errors are 90\% confidence intervals for one
interesting parameter ($\Delta\chi^2$ = 2.706). \par\noindent
$^{\rm (c)}$ this line flux gives an equivalent width EW = 29, 16 and 19 eV 
for SVP, A and D datasets respectively. \par\noindent
% $^{\rm (d)}$ \par 
}
\end{table}
%-----------------------------------------------------------------------------

Finally, as required by SVP--LECS data, we added to the model an
absorption edge at low energy\footnote{We refer to Grandi et al. 1997 for
a discussion of the origin of this feature.}, making the assumption that
it actually affects only the Seyfert--like continuum.

We then fit this model to the LECS--MECS--PDS data, for SVP and the AO1 
brightest and faintest states (A and D). 
The three states are fit {\it simultaneously}, i.e. the edge energy and
optical depth, the non--thermal power law index, and the LECS/MECS,
PDS/MECS inter--calibrations constants are forced to be the same for the
different sets of data in the fitting procedure.
Only the normalization of the non--thermal component is allowed to be
different for different datasets. 

The global fit is statistically good, with the best fit parameters
reported in Table~\ref{tab:fit_modellone}. 
The best fit model with all the different components relative to the SVP
and AO1 data is shown in Fig.~\ref{fig:models}, and the data to model
ratios for the three states are reported in Fig.~\ref{fig:3datasets}.

The resulting values of the inter--calibration constants (LECS/MECS =
$0.75 \pm 0.01$, PDS/MECS= $0.85 \pm 0.03$) are very close to the 
fiducial figures recommended by the \BS SDC. 

The edge is reasonably well constrained by the data at E$_{\rm
edge}=0.62\pm 0.08$ keV with an optical depth $\tau=1.7^{+1.3}_{-0.6}$.

The shape of the beamed component is determined by three parameters, 
namely the spectral index $\Gamma$, the energy above which the spectrum has  
a parabolic (in log) shape $E_{\rm break}$, and the $\nu f_{\nu}$ 
peak of the spectrum $E_{\rm peak}$. 

The beamed component turns out to have $\Gamma=1.54 \pm 0.01$,
fully consistent with spectral indices detected in the brightest states of
this source, when the non--thermal emission is likely to completely
dominate the X--ray flux.
In particular, we considered the recent results from ASCA by Cappi et
al. (1998).  The uncertainties in the calibration discussed by Cappi et
al. do not allow a quantitative comparison with the \BS
results (see e.g. the discussion about the iron line EW by these
authors). 
Nevertheless, we find that the model is also qualitatively 
consistent with the ASCA results, in particular with the hardening of
the 2--10 keV spectrum as the source gets brighter.

The curved shape of the non--thermal continuum nicely accounts for the
observed steep spectral index in the PDS range (see Fig.~\ref{fig:indices}).

The quality of the fit is rather insensitive to the energy of the peak of
the parabolic component, and good fits are compatible with the power law
starting steepening at energies above the MECS range and peaking around a few 
MeV. The values reported in Table~\ref{tab:fit_modellone} are obtained with
the peak of the parabolic component fixed at 2 MeV.

Our results show that \BS observations of 3C 273 can be interpreted in a
framework where the increase of the non--thermal emission in the AO1--A
swamps out the spectral features connected to the Seyfert--like component,
such as the soft-excess, the absorption edge and the iron line.
Moreover, it should be stressed that, in this model, the steep `primary'
thermal emission in the SVP state exceeds the non--thermal one at the
lowest energies and naturally reproduces the steeper power law detected 
in the LECS energy range. 
The steepening of the higher energy spectrum observed in the AO1
data is simply interpreted as the curvature of the inverse Compton
emission reaching its peak.

Once again, we stress that the model has not been constructed in order
to best fit all the available data. The number of free parameters
coupled with the data uncertainties would make this meaningless. Note
in particular that there is no reason why any thermal component should
be constant and there is actually clear evidence for variability in
the spectral slope (e.g., observation C in AO1), with no
apparent correlation with the source intensity. Similar uncorrelated 
spectral variations have been also reported in EXOSAT and GINGA 
observations (Turner et al. 1990). Also, we can not exclude that the 
reflection hump is at a level lower than assumed (i.e., $\Omega/2\pi<1$), 
as seems the case in Broad Emission Line Galaxies (Wozniak et al. 1998).

However, from the fitting of the three states (SVP, low and high
states of AO1) we conclude that Seyfert--like emission from an
accretion disk and its corona is likely present, roughly at comparable
flux level with the highly variable beamed non--thermal emission from
a relativistic jet.  The model in fact well reproduces three
\BS observations and is consistent with the ASCA ones, by only allowing
for a change in the normalization of the non--thermal emission.

                   %---------------------%

\section{Summary and conclusions}

The results of \BS observations of 3C 273 performed in Jan. 1997 have been
presented.  The source varied within the observations and with respect
to previous SVP data. Most notable is the variation of X--ray features
clearly detected during the low state SVP, and barely present in the
AO1 data (and qualitatively consistent with the ASCA results 
recently reported by Cappi et al., 1998).

The variable strength of the observed spectral features is consistent
with a spectrum consisting of X--ray emission by a typical Seyfert--like
nucleus, namely direct and reprocessed radiation dissipated in or
above an optically thick accretion disk, and a variable non--thermal
beamed component generated by the plasma moving in the relativistic
jet.  Although the AO1 data could be satisfactorily modelled by a
single power law, slightly steepening above $\sim 10$ keV, the
comparison with the SVP phase suggests that any thermal emission is
indeed more evident in low source states (corresponding to weaker
non--thermal radiation).  Indeed the data available can be basically
accounted for in this picture, by varying the absolute value of the
normalization of the non--thermal emission. Clearly, monitoring of 3C
273 to derive variability information on the broad \BS energy range
will allow to develop a more consistent and detailed model.

Indications of thermal/Seyfert--like emission in 3C 273 have been
recognized for a long time, in different spectral bands, showing that
in this source both thermal and non--thermal emission processes are taking
place, with comparable radiative dissipation rate.  The X--ray results
fit well and support this picture. Further observations of the high
energy broad band spectra of radio--loud objects are critical in
understanding the/any relationship between the dissipation in the
accreting and ejected material.

\begin{acknowledgements}
We thank the SAX--Team for providing valuable help and support, and 
M. Cappi, M. Guainazzi and S. Molendi for helpful comments and useful 
discussions. This research, in its early stages, made use of SAXDAS linearized
and cleaned event files (rev0.0) produced at the \BS Science Data Center. 
GF and AC acknowledge the Italian MURST for financial support.
\end{acknowledgements}

%\newpage


\begin{thebibliography}{}
\bibitem[1979]{bessel79}{Bessell M.S., 1979, PASP 91, 589}
\bibitem[1997]{boella97}{Boella G. et al., 1997, A\&AS, 122, 299}
\bibitem[1970]{bowyer70}{Bowyer C.S., Lampton M., Mack J., de Mendonca F.,
1970, ApJ, 161, L1} 
\bibitem[1996]{cappi96}{Cappi M., Matsuoka M., 1996, in Proc. of 2nd
Integral Workshop, ``The Transparent Universe", St Malo, Ed. C. Winkler,
T.J.-L Courvoisier \& P. Durouchoux, ESA SP-382, p389 }
\bibitem[1998]{cappi98} {Cappi M., Matsuoka M., Otani M., Leighly K.M., 1998, PASJ in press}
\bibitem[1995]{chakra95} {Chakrabarti S.K., Titarchuk L.G., 1995, ApJ, 455, 623}
\bibitem[1987]{courvo87} {Courvoisier T.J.-L., et al., 1987, A\&A, 176, 197}
\bibitem[1997]{geddon}{Fossati G., Haardt F., 1997, SISSA/ISAS report, {\tt
ref.~SISSA~146/97/A}, (August 1997)}
%\bibitem[1997]{frontera97}{Frontera F., Costa E., Dal Fiume D., Feroci M., Nicastro L., Orlandini M., Palazzi E., Zavattini G., 1997, A\&AS, 122, 357}
\bibitem[1996]{gmd96} {Ghisellini G., Maraschi L., Dondi L., 1996, A\&AS, 120, 503} 
\bibitem[1997]{grandi273} {Grandi P., et al., 1997, A\&A, 325, L17}
\bibitem[1997]{Guainazzi97} {Guainazzi M., 1997, private communication}
\bibitem[1997]{GuainazziGrandi97} {Guainazzi M., Grandi P., 1997, \BS
SDC technical report TR 14, August 1997}
\bibitem[1991]{HM91} {Haardt F., Maraschi L., 1991, ApJ, 380, L51}
\bibitem[1997]{lincei273} {Haardt F., et al. 1997, In:  ``The Active X--ray
Sky: Results from \BS\ and Rossi--XTE", Scarsi L., Bradt H., Giommi P., 
Fiore F. (eds.), Nuclear Physics B Proc. Supp., in press}
\bibitem[1992]{harmon92} {Harmon B.A., et al., 1992, The Compton Observatory 
Science Workshop, NASA CP3137, eds C.R. Shrader, N. Gehrels, B. Dennis, 69}
\bibitem[1992]{hartman92} {Hartman R.C. et al., 1992, ApJ, 385, L1}
\bibitem[1993]{hermsen93} {Hermsen W. et al., 1993, A\&A, 97, 97}
\bibitem[1995]{johnson95} {Johnson W.N. et al., 1995, ApJ, 445, 182} 
\bibitem[1995]{leach95} {Leach C.M., McHardy I.M., Papadakis I.E., 1995, MNRAS, 272, 221}
\bibitem[1995]{lockman95} {Lockman F.J., Savage B.D., 1995, ApJS, 97, 1}
\bibitem[1993]{mannheim93} {Mannhein K., 1993, A\&A, 269, 67}
\bibitem[1997]{gifco273} {Maraschi L., Fossati G., Haardt F., 1997, 
Conf. Proc. S.I.F., vol. 58, p35} 
\bibitem[]{} {Marscher A., Travis J.P., 1996, A\&AS, 120, 537}
\bibitem[]{} {McNaron--Brown K., et al., 1995, ApJ, 451, 575}
\bibitem[]{} {Nandra K., Pounds K.A., 1994, MNRAS, 268, 405}
\bibitem[]{} {Parmar A.N., et al., 1997, A\&AS, 122, 309}
\bibitem[]{} {Ramos E., Kafatos M., Fruscione A., Bruhweiler F.C., McHardy 
I.M., Hartman R.C., Titarchuk L.G., von Montigny C., 1997, ApJ, 482, 167}
\bibitem[]{} {Shapiro S.L., Lightman A.P., Eardley D.M., 1976, ApJ, 204, 187}
\bibitem[]{} {Smith P.S., Balonek T.J., Heckert P.A., Elston R., 
Schmidt G.D., 1985, AJ 90, 1184}
\bibitem[]{} {Staubert R., 1992, in X--Ray Emission from AGN and the Cosmic 
X--Ray Background, ed W. Brinkmann \& J. Tr\"umper (MPE Rep. 235), p42}
\bibitem[]{} {Turner M.J.L., Courvoisier T., Staubert R., Molteni D., 
Tr\"umper J., 1985, in Proc. of 18th ESLAB Symp., ed A. Peacock 
(Dodrecht: Reidel), p623}
\bibitem[]{} {Turner M.J.L., et al., 1990, MNRAS, 244, 310}
\bibitem[]{} {Turner M.J.L., Weaver K.A., Mushotzky R.F., Holt S.S., 
Madejski G.M., 1991, ApJ, 381, 85}
\bibitem[]{} {von Montigny C., et al., 1997, ApJ, 483, 161}
\bibitem[]{} {Williams O.R., et al., 1992, ApJ, 398, 157}
\bibitem[]{} {Wozniak P.R., Zdziarski A.A., Smith D., Madejski G.M., Johnson 
W.N., 1998, MNRAS, in press}
\bibitem[]{} {Yaqoob T., et al., 1994, PASJ, 46, L49}
\end{thebibliography}
\end{document}